\documentclass{ws-procs975x65}
\usepackage{epsfig}
\usepackage{sidecap}

\begin{document}
\title{Non-adiabatic effects in the irradiation of ethylene}

\author{Z.P. Wang,$^{1,2,3,4,5}$ P.M. Dinh,$^{4,5}$\footnote{Corresponding
author. Email: dinh@irsamc.ups-tlse.fr} P.-G. Reinhard,${^6}$ E. Suraud,$^{4,5}$ and F.S. Zhang$^{2,3}$}
\address{$^1$School of Science, JiangNan University, Wuxi 214122, China\\
$^2$The Key Laboratory of Beam Technology and Material
Modification of Ministry of Education, College of Nuclear Science
and Technology, Beijing Normal University,
Beijing 100875, People's Republic of China\\
$^3$Beijing Radiation Center, Beijing 100875, China\\
$^4$Universit\'{e} de Toulouse; UPS; Laboratoire Physique Th\'eorique (IRSAMC), F-31062 Toulouse, France\\
$^5$CNRS; LPT(IRSAMC), F-31062 Toulouse, France\\
$^6$Institut f\"{u}r Theoretische Physik, Universit\"{a}t
Erlangen, Staudtstrasse 7, D-91058 Erlangen, Germany}
\date{\today }

\begin{abstract}
In the framework of the time dependent local density approximation,
applied to valence electrons, coupled non-adiabatically to molecular
dynamics of ions, the irradiations of ethylene by laser and fast
charged projectiles are studied. We find that the Coulomb
fragmentation sensitively depends on the laser frequency and on the 
charge of the projectile.
\end{abstract}

\section{Introduction}
\label{intro}

It is well know that the interaction of ionizing radiation with
biological tissue can induce severe damage to DNA \cite{1,2}.
%
With the development of ion sources such as the
electron cyclotron resonance (ECR) ion source and modern
accelerators such as cooling storage ring (CSR), a lot of
investigations have been devoted to understanding the molecular
mechanisms underlying biological radiation damage. More recently, a
number of experimental investigations have been addressed concerning  the
action of ions on molecules with high and low energies
\cite{3,4,5,6,7,8,Sayler,H.Luna,Rentenier,Melo}. From the mass
spectra, the relative cross sections for the different ionization
and fragmentation channels can be evaluated. One can observe not
only the process of electron capture but also the highly excited
processes, such as electronic emission and Coulomb explosion due to the
induced ionization.

Impact of radiation also includes excitation by photons
and we will discuss their effects as well.
The advantage of photons is that modern lasers provide flexible and
powerful pulses for studies of photo-induced dynamics, such as optical
response \cite{U.Kreibig}, multi-photon ionization \cite{Faisal}, and
Coulomb explosion of clusters \cite{J.Zweiback}, high-order harmonic
generation \cite{J.Itatani, P.B.Corkum, P.Agostini}, bond softening
\cite{G.Yao, SK}, and charge resonance enhanced ionization
\cite{T.Zuo, T.Seideman}.

The perfect theoretical description of these mechanisms 
would require to solve
the time-dependent Schr\"{o}dinger equation for all electrons and
all nuclear degrees of freedom. However, this ultimate attack
exists only for very small systems, such as atoms
\cite{JPHansen, Parker}, H$_2$ and H$_2^+$ \cite{Chelkowski,
Kreibich, ADBandrauk, XBB, Liang}. For larger systems,
more approximate ways have to be found.
Recently, Horsfield \emph{et al.} used correlated electron-ion
dynamics to study the excitation of atomic motion by energetic
electrons \cite{Horsfield}. Saalmann \emph{et al.} have developed a
non-adiabatic quantum molecular dynamics (NA-QMD) to study different
non-adiabatic processes in different systems \cite{Saalmann,
T.Kunert, Kunert} and Calvo \emph{et al.} used a combined method to
study the fragmentation of rare-gas clusters \cite{F.Calvo}. It is also
interesting to develop a fully-fledged coupled ionic and electronic dynamics.
At the side of electronic dynamics, the well tested
Time-Dependent Density Functional Theory at the level of the
Time-Dependent Local Density Approximation (TDLDA) \cite{NATO} is
used. The ions are treated by classical molecular dynamics (MD). This
altogether provides a coupled TDLDA-MD, for a review see
\cite{F.Calvayrac}. Up to now, this approach has been applied to study the
dynamical scenarios of simple metal and hydrogen clusters
\cite{A.Castro,D.Dundas,E.Suraud,F.Cal,L.M.Ma,M.Ma,F.S.}. Besides
these systems, organic molecules are particularly interesting cases and recently 
motivated a lot of theoretical works
to investigate either the optical response in the
linear domain \cite{Takashi} or the dynamics of it without and with
considering the external laser field \cite{Ben-Nun, Kunert}.

In this paper, the irradiations of ethylene by laser and fast
charged projectiles are studied by using a coupled TDLDA-MD
\cite{F.Calvayrac} and various non-adiabatic effects are
investigated. The paper is organized as follows. Section 2 provides
a short presentation of the theoretical and numerical approach.
Section 3 first gives the optical response properties of ethylene
and the effect of ionic motion on the excitation dynamics is
discussed. Then different excitation scenarios of ethylene as a
function of the laser frequency and charged projectiles are
presented. Section 4 gives some conclusions.

\section{Theory}
\label{sec:1}

In this section, we briefly  represent the real-time method in TDLDA-MD. The
molecule is described as a system composed of valence electrons and
ions. Our test case, the ethylene molecule C$_2$H$_4$, has 12 valence
electrons and 6 ions. The interaction between ions and electrons is
described by means of norm-conserving pseudopotentials
\cite{Goedecker}.

Valence electrons are treated by TDLDA, augmented with an
average-density self-interaction correction (ADSIC)
\cite{Legrand}. They are represented by single-particle orbitals
$\phi_{j}(\textbf{r},t)$ satisfying the time-dependent Kohn-Sham
(TDKS) equations \cite{Gross},
\begin{eqnarray}
i\frac{\partial}{\partial{t}}\phi_{j}(\textbf{r},t) & = & \
\hat{H}_\mathrm{Ks}\phi_{j}(\textbf{r},t) 
=
\left(-\frac{\nabla^{2}}{2 m_{\rm el}}+V_\mathrm{eff}(\textbf{r},t)\right)
\phi_{j}(\textbf{r},t),
\quad,\quad 
j=1,...,N.
\end{eqnarray}
The term $V_\mathrm{eff}$ stands for the Kohn-Sham effective potential and is
composed of four parts,
\begin{eqnarray}
V_\mathrm{eff}(\textbf{r},t)) 
= 
V_\mathrm{ion}(\textbf{r},t)+V_\mathrm{ext}(\textbf{r},t)
+V_\mathrm{H}[n](\textbf{r},t)+V_\mathrm{xc}[n](\textbf{r},t),
\end{eqnarray}
where $V_\mathrm{ion} = \sum_{I}V_{ps}(\textbf{r}-\textbf{R}_I)$ is the ionic
background potential, $V_\mathrm{ext}$ the external potential, $V_\mathrm{H}$
stands for the time-dependent Hartree part and $V_\mathrm{xc}$ is the
exchange-correlation (xc) potential. The electron density is given
by $n(\textbf{r},t) = \sum_{j}|\phi_{j}(\textbf{r},t)|^{2}$.
The xc potential $V_\mathrm{xc}[n](\textbf{r},t)$ is a functional of the
time-dependent density and has to be approximated in practice. The
simplest choice consists in the TDLDA, defined as
\begin{eqnarray}
V_\mathrm{xc}^{\rm TDLDA}[n](\textbf{r},t) =
\left.
\frac{\delta \epsilon_\mathrm{xc}^{\rm hom}[n]}
{\delta n}
\right|_{n=n(\textbf{r},t)},
\end{eqnarray}
where $\epsilon_\mathrm{xc}^{\rm hom}[n]$ is the xc energy density of the
homogeneous electron gas. For $\epsilon_\mathrm{xc}^{hom}$ we use the
parameterization of Perdew and Wang \cite{Perdew}.
The form of the pseudopotential for a covalent molecule is taken from
\cite{Goedecker}, including a non-local part. The TDLDA approximation
is augmented by an Average Density Self-Interaction Correction
(ADSIC)~\cite{Legrand} to put the single-particle energies at their
correct values.

The ground state wave functions are determined by the damped
gradient method \cite{Calvayrac}. The TDLDA equations are solved
numerically by time-splitting technique \cite{F.C}. The
nonlocal part contained in $\rm V_{\rm eff}$ is dealt in an
additional propagator and treated with a third-order Taylor
expansion of the exponential \cite{Ann}. Absorbing boundary
conditions are employed to avoid reflecting electrons
\cite{Ullrich}.

Ions are treated classically and propagated by classical 
Molecular Dynamics (MD) equations
\begin{eqnarray}
m_{I}\frac{\rm d^{2}\textbf{R}_{I}}{\rm dt^{2}}=\textbf{F}_{I}(\textbf{R}_{I},t)
\end{eqnarray}

The external perturbation by the laser field, neglecting the magnetic
field component, acting both on electrons and ions, is given as 
\begin{eqnarray}
V_\mathrm{ext,las} = E_{0}zf_\mathrm{las}(t)\cos(\omega_\mathrm{las}t)
\end{eqnarray}
where $E_{0} \propto {\sqrt{I}}$, $I$ denoting the intensity, $z$ is
the dipole operator, $f_\mathrm{las}(t)$ is the pulse profile and
$\omega_\mathrm{las}$ is the laser frequency. In this paper,
$f_\mathrm{las}(t)$ is chosen as a $\cos^{2}$ in time.

The external perturbation by a charged projectile is modeled as
\begin{eqnarray}
V_\mathrm{ext,ion} = - \frac{Qe^2}
{\sqrt{\left[x-(\upsilon_{proj}t+x_0)\right]^2+y^2+(z-b)^2}}
\end{eqnarray}
where $Q$, $\upsilon_\mathrm{proj}$, $x_0$ denote the charge, the
velocity and the initial position of the projectile respectively; 
$b$ is the
impact parameter. The excitation amplitude may be tuned through
variation of $Q$, $\upsilon_\mathrm{proj}$, and $b$. Here we focus on
the influence of $Q$, $b$ and collision orientation. The velocity
is fixed to $\upsilon_\mathrm{proj}=20\, a_0$/fs.

The calculations are carried out using a cubic box of size $72\times 
72\times 64$
with a grid spacing of 0.41~$a_{0}$ and a constant time step $\Delta
t=6\times 10^{-4}$~fs. The dipole moment can be obtained by $D_i(t)=
\int \textrm d^{3}\mathbf r \, r_{i}\, n(\textbf{r},t)$, with $i=x,y,z$. 
The number of
escaped electrons is defined as $N_{\mathrm{esc}} = N(t=0) -
\int_{V} \textrm d^{3}\mathbf r \, n(\textbf{r},t)$, where $V$ is a 
sufficiently large volume surrounding
molecule. A detailed link with experiments is the probabilities
$P^k(t)$ of finding the excited molecule in one of the possible
charge states $k$ to which they can ionize. The formula are deduced
from the occupied numbers of the time-dependent single-particle
orbitals and are taken from \cite{Ullrich}.

\section{Results and discussion}

\subsection{Basic properties of C$_2$H$_4$}

\begin{SCfigure}[0.7][htbp]
\includegraphics[width=9cm,angle=0]{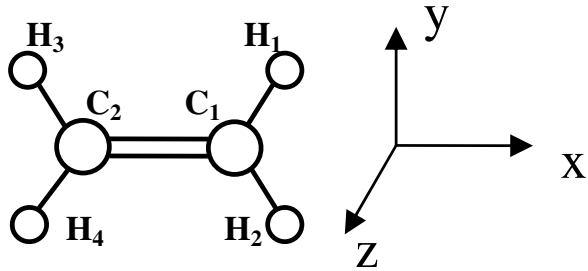}
\caption{Ionic structure of ethylene.}
\label{structure}       
\end{SCfigure}

Fig.~\ref{structure} shows the spatial structure of ethylene. 
The molecule is planar and we place it
in the $x$-$y$ plane with two carbon atoms on the $x$ axis and
four hydrogen atoms in $x$-$y$ plane. The center of mass is at the
origin. The single-electron energies range from $-23.3$ eV for
the deepest bound valence level to the HOMO at $-11.5$ eV. The latter
is in good agreement with the experimental ionization potential (IP)
of 11 eV~\cite{NIST} (relative error of 4.5~\%). Similar 
errors are obtained when comparing the C-C and the C-H bond lengths
(2.43 and 1.97 $a_0$ respectively, while the experimental values are
2.53 and 2.05 $a_0$~\cite{NIST}).

\begin{SCfigure}[0.7][htbp]
\includegraphics[width=9cm,angle=0]{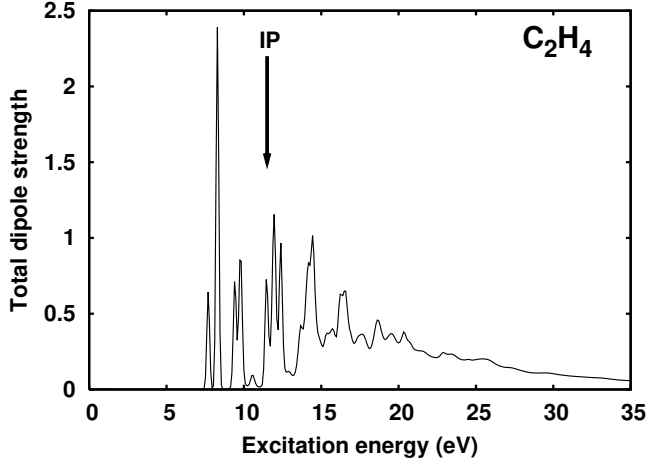}
\caption{Total photo-absorption strength of C$_2$H$_4$. The vertical
  arrow corresponds to the value of the ionization potential of the molecule.}
\label{opt_resp}       
\end{SCfigure}

Fig.~\ref{opt_resp} shows the optical response, i.e. the
photo-absorption strength distribution, of C$_2$H$_4$. 
It shows
strong, isolated peaks at lower energies and changes to a continuum
above the IP (11.5 eV).  This result is consistent with former TDLDA
calculations~\cite{Takashi}. The lowest peak is at 7.74 eV, in good
agreement with previous CI calculations (7.76
eV~\cite{C.Petrongolo}). We obtain the highest peak in the optical
response lying at 8.16 eV. The frequency selective laser pulses
will be sensitive to this much fluctuating spectrum while a collision
with fast ions will deliver a broad spectrum of frequencies and will 
excite all modes at once.

\subsection{Laser excitation}

\subsubsection{The key role of non-adiabatic couplings}

\begin{SCfigure}[0.8][htbp]
\includegraphics[width=8cm,angle=0]{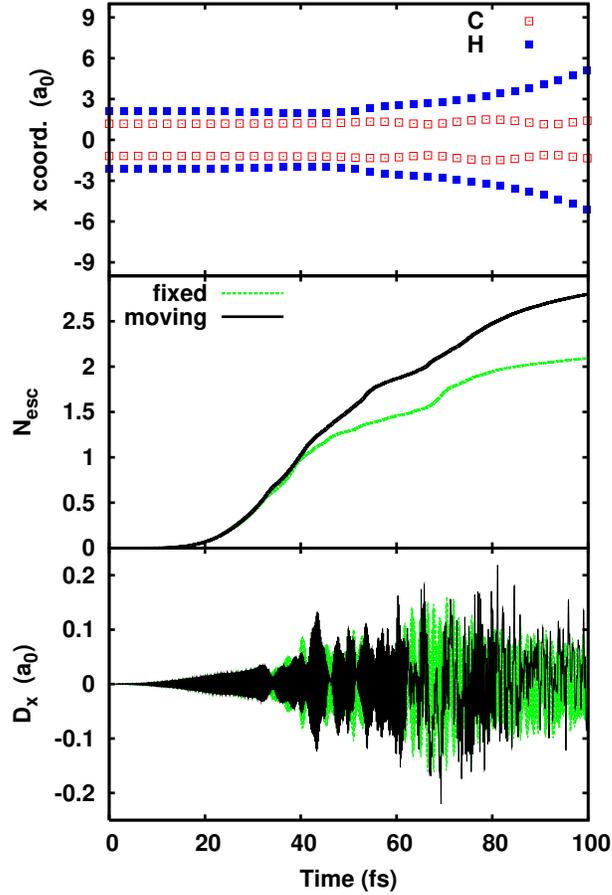}
\caption{Excitation of ethylene induced by a
laser pulse ($I=10^{13}$~W/cm$^2$, $\omega_\mathrm{las}=8.16$~eV and
FWHM of 100~fs), for fixed or moving ions. Top panel: time evolution
of the $x$ ionic coordinates, when ionic motion is allowed. Middle and
bottom:
$N_{\rm esc}$ and dipole moment $D_x$ as a function of time, for
moving (dark lines) or fixed (green curves) ions.}
\label{fig:ionic}
\end{SCfigure}

To investigate the effect of the motion of the ionic centers of the
molecule on the excitation dynamics, we compare, for the case of a laser
excitation, calculations either with full ionic motion or with fixed
ionic configuration. The laser intensity is $10^{13}$ W/cm$^2$ and its full 
width at half maximum (FWHM) is 100 fs. We have chosen the laser 
frequency equal to 8.16 eV, that is 
the value of the highest peak in the optical response (see Fig.~\ref{opt_resp}). We 
thus expect a resonant excitation of the ethylene.
Fig.~\ref{fig:ionic} shows the time
evolution of the $x$ ionic coordinates, the number of escaped electrons
($N_\mathrm{esc}$), and the dipole moment in the $x$ direction, $D_x$,
for moving and fixed ions.  
Up to about 35 fs, no significant difference is visible in $D_x$ and
$N_{\rm esc}$. From then on, comparing both cases, the dipole moments
diverge rapidly in time and more complex pattern are observed for the
case of moving ions. Consequently, the total ionization also
differs and more electronic emission is obtained when ions can move
(0.8 additional electron is extracted).
The ions precisely start to exhibit a sizeable motion at about 40
fs (see top panel). The C-C bond oscillates (see the open squares)
while the H (close squares) escape the system, due to the 
large charge. The non-adiabatic coupling between ions and electrons
allows a feed-back of the ionic motion on the electronic response, as
can be tracked on the dipole oscillations.
The enhancement of ionization for the moving ions case
is related to the expansion of the molecule.
Indeed, after increasing ionization, the optical spectrum is blue-shifted, 
so $\omega_\mathrm{las}$ becomes smaller than the resonant frequency.
However, the Coulomb force from increased net charge impels the
expansion of ions which red-shifts of the optical spectrum
and allows $\omega_\mathrm{las}$ to be in the region of resonance again.
The difference of electronic emission above shows the importance of
treating electronic and ionic dynamics simultaneously. 

\subsubsection{Scanning laser frequency}

\begin{SCfigure}[0.7][htbp]
\includegraphics[width=10cm,angle=0]{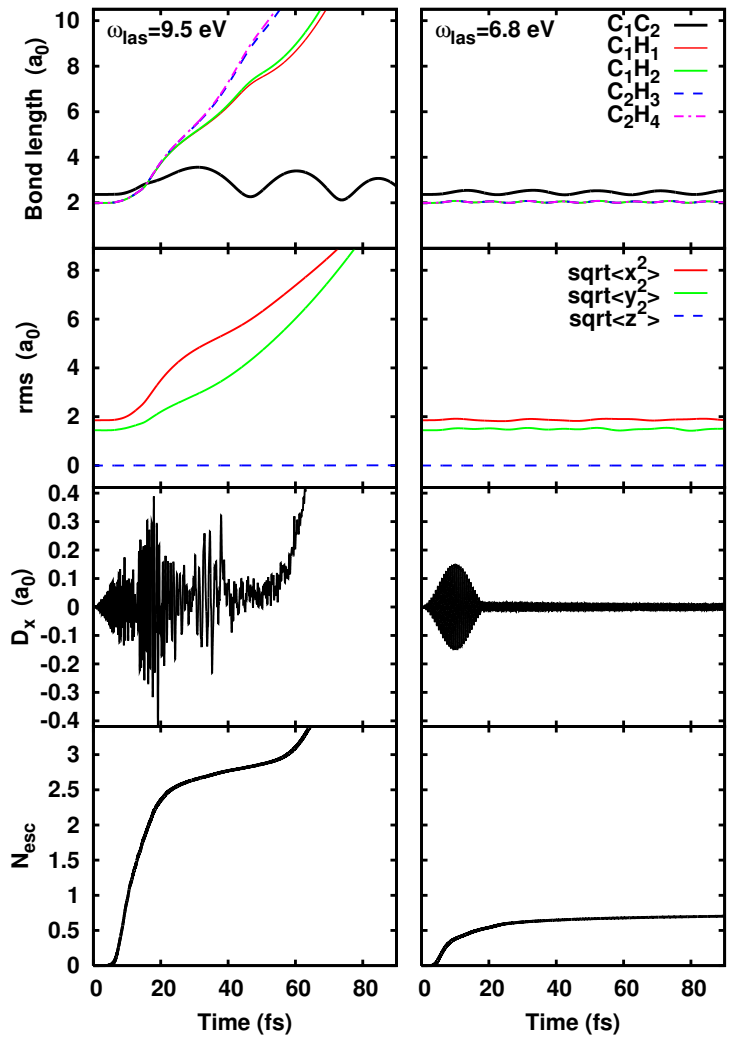}
\caption{Excitation of ethylene by a laser pulse of $I=10^{14}$~W/cm$^2$,
FWHM of 20~fs, polarization along $x$ axis,
for two different frequencies, $\omega_\mathrm{las}=9.5$~eV 
(left column), and $\omega_\mathrm{las}=6.8$~eV (right column). 
From top to bottom panels: time evolution of
bond lengths, extensions of the ionic distribution in three
directions, dipole signal along the laser polarization and the
number of escaped electrons.} \label{fig:freq}
\end{SCfigure}

To explore in more detail the influence of the laser frequency, we consider two
cases, namely $\omega_\mathrm{laser}=9.5$~eV, lying
is in the resonant region, and $\omega_\mathrm{laser}=6.8$~eV
which is off-resonant. Both cases are depicted in Fig.~\ref{fig:freq} in
the left and the right columns respectively.
The laser parameters are a polarization along $x$ axis, an intensity
of $I=10^{14}$~W/cm$^2$, and a rather short FWHM of 20~fs.
The bond lengths and the r.m.s. radii in the resonant case, top-left
panels of 
Fig.~\ref{fig:freq}, show much different dynamics for the various 
bonds.  The C-H bonds grow quickly and monotonously, which
means that the bonds are broken and hydrogen ions are flying apart in
the $x-y$ plane.
The C-C bond (dark solid line), on the contrary, oscillates with a
rather large amplitude during
the whole time span. The off-resonant case, top-right panels of
Fig.~\ref{fig:freq}, shows regular and faint oscillations for all
bonds and r.m.s radii. The molecule remains unfragmented, although
highly excited.

The ionization $N_{\mathrm esc}$ and the dipole evolution shown in 
the bottom panels of Fig.~\ref{fig:freq}, corroborate these findings. 
In the off-resonant case (right panels), the dipole nicely follows the
laser field, producing an emission of 0.6-0.7 electron during the
pulse duration. As soon as the laser is  
switched off, faint dipole oscillations remain and $N_{\rm esc}$ levels off. 
The resonant case visibly produces a complex pattern in the dipole evolution
with large amplitudes; more ionization is observed, driving the molecule above the
limits of binding.  Note that $N_\mathrm{esc}$ starts to increase at around 5
fs while the ionic motion (see the bond lengths) follows later.  Ionization is
directly caused by the laser pulse and electrons react quickly because
they have a low mass. The ions react indirectly, namely to the
Coulomb pressure induced by ionization, and more slowly due to their
larger mass.

The above discussion demonstrates the versatility of a laser excitation
when playing with its frequency (at low intensities). Indeed the value
of this frequency plays an important role in the excitation dynamics of
ethylene. In the resonant region, enhanced ionization often leads to a Coulomb 
fragmentation of the molecule.

\subsection{An example of ionic collision}

\begin{SCfigure}[0.7][htbp]
\includegraphics[width=7cm,angle=0]{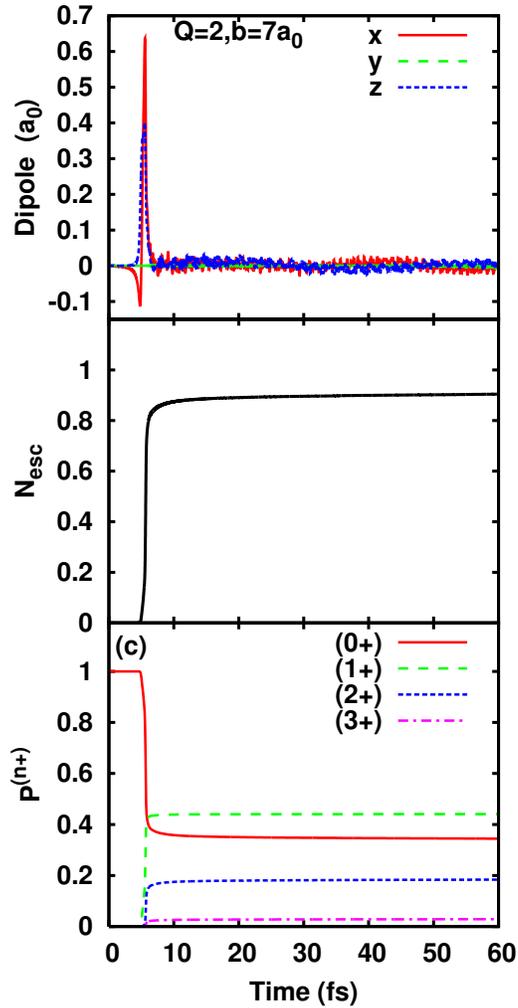}
\caption{Time evolution of the dipole moments (top), of the total
  number of escaped electrons (middle) and ionization probabilities
  (bottom), after collision of C$_2$H$_4$ with a projectile of charge
  $Q=2$, impact parameter $b=7 a_0$ and velocity along the $x$
  direction of $\upsilon_\mathrm{proj}=20$~$a_0$/fs.} \label{example}
\end{SCfigure}

We finally end with a typical example of collision between
ethylene and a projectile with charge $Q=2$.  The projectile moves
along $x$ direction at the velocity of 20~$a_0$/fs,
with an offset of $b=7$~$a_0$ in $z$ direction.
The initial position of the projectile is ($-100\, a_0$, 0, $7\,a_0$), thus
safely far away from the ethylene.  
Fig.~\ref{example} displays the time evolution of the dipole moments
in the three spatial directions, $N_{\rm esc}$ and the ionization
probabilities $P^{(n+)}$.
Let us start with the dipole response after collision with the
projectile (top panel). Since the projectile
moves in the $x-z$ plane along the 
$x$ axis, no significant excitation along the $y$ direction is
expected. And indeed, the dipole moment in that direction, $D_y$, is
vanishingly small. On the contrary, $D_x$ and $D_z$ exhibit at 
time of closest approach (at about 5 fs) an almost instantaneous large
shift and quickly relaxe towards much more gentle oscillations that
last for a long time. Note that the shift in $x$ direction is higher
than that in $z$ direction. This is consistent with the fact that the
extension of the electronic cloud is larger in $x$ than in $z$
directions (see the ethylene configuration in Fig.~\ref{structure}).
The total ionization, shown in the middle panel of Fig.~\ref{example},
evolves according to the dipole excitation~: it jumps almost
immediately at impact time in correlation to the huge dipole moments,
while the faint later dipole oscillations do not induce further
electron emission.
The bottom panel of Fig.~\ref{example} shows the detailed
ionization probabilities. Their time evolution proceed similarly to
that of $N_{\rm esc}$. Indeed $P^{(0+)}$ falls down abruptly at the
time of closest approach, while the other charge state probabilities
grow very quickly. Then all probabilities level off and no evolution
in time is observed. The charge state $1^+$ has the largest
probability in accordance with a net ionization of nearly 1 (see
middle panel), but a sizeable value of $P^{2+}$ of 20~\% is
nevertheless attained. As for higher charge states, their
probabilities are negligible as expected.

At the side of the ions, this fast collision only induces a visible
motion in the $x-y$ plane. More precisely, faint
oscillations parallel to the $x$ axis of the C-C bond (aligned with 
the $x$ axis) are observed, while the C-H bond exhibit larger (but
still small) oscillations, because of the smaller mass of the H
compared with that of the C. In the $y$ direction, the C atoms simply
do not move and the H gently oscillate around their equilibrium
positions. 

Apart from the different dipole responses, this collision
case is close to the off-resonant laser excitation presented in
Fig.~\ref{fig:freq}, right column. Indeed, the total ionization there
is quite similar (0.7 compared with 0.9 here). However, the ethylene
molecule seems able to cope such a ionization and remains in one piece.

Finally, we just mention here some other cases of projectile collision
that we have explored. Increasing the charge of the projectile can
cause a direct Coulomb fragmentation of the ethylene molecule.
We have also varied the value and the direction of the impact
parameter $b$. When the projectile moves perpendicular to the ethylene
plane, the electronic response is of course weaker, since the
interaction time is shorter than in the case of a projectile moving
parallel to the molecular plane. Still the investifated test cases 
lead us to conclude that the dominant (first order) parameter is the 
actual (absolute) value of $b$. 


\section{Conclusions}
In this paper, we have demonstrated the capability of TDLDA-MD to
describe complex coupled ionic and electronic excitations. 
We discussed various excitation scenarios of ethylene
subjected to different laser pulses and fast charged
projectiles. These scenarios involve both electrons and 
ions but the relative role of each species does depend on the actual
excitation conditions of the laser pulse and charged particles. The
response in time of the electron cloud, such as dipole deformation, total
number of escaped electrons and ionization probabilities have been
presented. Varying the laser frequency with fixed laser intensity and
pulse length indicates that the appearance of
de-excitation$/$explosion scenarios depends on the relationship of the
laser frequency with the eigenfrequencies of the system. We have also
explored the case of a projectile collision at moderate velocity, 
a situation that can be easily handled experimentally. A very
fast excitation is reported. A systematic study of the dependence
on charge, velocity and impact parameter of the projectile will be
presented in a forthcoming publication. 

\section*{ACKNOWLEDGEMENTS}
This work was supported by the National Natural Science Foundation
of China (Grants No. 10575012 and No. 10435020), the National Basic
Research Program of China (Grant No. 2010CB832903), the Doctoral
Station Foundation of Ministry of Education of China (Grant No.
200800270017), the scholarship program of China Scholarship Council
and the French Agence Nationale pour la Recherche
(ANR-06-BLAN-0319-02). Calculations have been performed on the French
computational facilities CalMiP, CINES and IDRIS.

\end{document}